\documentclass[useAMS,usenatbib]{mn2e}
\usepackage{mathrsfs}
\usepackage{graphicx,amssym,color}
\newcommand{\beq}{\begin{equation}}
\newcommand{\eeq}{\end{equation}}
\newcommand{\beqa}{\begin{eqnarray}}
\newcommand{\eeqa}{\end{eqnarray}}

\usepackage{color}
\def\gsim { \lower .75ex \hbox{$\sim$} \llap{\raise .27ex \hbox{$>$}} }
\def\lsim { \lower .75ex \hbox{$\sim$} \llap{\raise .27ex \hbox{$<$}} }
\def\proptosim { \lower .75ex \hbox{$\sim$} \llap{\raise .27ex \hbox{$\propto$}} }

\newcommand{\bc}{\begin{center}}
\newcommand{\ec}{\end{center}}

\newcommand{\Msun}           {\,{\rm M}_\odot}

\newcommand{\kms}            {\,\,{\rm km}\,\,{\rm s}^{-1}}
\newcommand{\kmsMpc}         {\,\,{\rm km}\,\,{\rm s}^{-1}\,{\rm Mpc}^{-1}}

\title[The APOSTLE project] {The APOSTLE project: Local Group
  kinematic mass constraints and simulation candidate selection}
\author[Fattahi et al. ]{
\parbox[t]{\textwidth}{
       Azadeh Fattahi$^{1}$\thanks{Email: azadehf@uvic.ca}, Julio
         F. Navarro$^{1,2}$, Till Sawala$^{3}$ , Carlos S. Frenk$^{3}$ ,
         Kyle A. Oman$^{1}$ , Robert A. Crain$^{4}$ , 
         Michelle Furlong$^{3}$ ,  Matthieu Schaller$^{3}$ , Joop Schaye$^{5}$,
          Tom Theuns$^{3}$, Adrian Jenkins$^3$
}
  \\ \\
\parbox[t]{\textwidth}{
$^1$Department of Physics and Astronomy, University of Victoria, PO
Box 3055 STN CSC, Victoria, BC, V8W 3P6, Canada\\
$^2$Senior CIfAR Fellow.\\
$^{3}$ Institute for Computational Cosmology, Department of Physics, University of Durham, South Road, Durham DH1 3LE, United Kingdom\\
$^{4}$ Astrophysics Research Institute, Liverpool John Moores University, IC2, Liverpool Science Park, 146 Brownlow Hill, Liverpool, L3 5RF, United Kingdom\\
$^{5}$ Leiden Observatory, Leiden University, PO Box 9513, NL-2300 RA Leiden, the Netherlands\\
}
}

\begin{document}

\date{\today}

\pagerange{\pageref{firstpage}--\pageref{lastpage}}
\pubyear{2015}

\maketitle

\label{firstpage}

\begin{abstract}
  We use a large sample of isolated dark matter halo pairs drawn from
  cosmological N-body simulations to identify candidate systems whose
  kinematics match that of the Local Group of Galaxies (LG). We find,
  in agreement with the ``timing argument'' and earlier work, that the
  separation and approach velocity of the Milky Way (MW) and Andromeda
  (M31) galaxies favour a total mass for the pair of
  $\sim 5\times 10^{12} \Msun$. A mass this large, however, is
  difficult to reconcile with the small relative tangential velocity
  of the pair, as well as with the small deceleration from the Hubble
  flow observed for the most distant LG members. Halo pairs that match
  these three criteria have average masses a factor of $\sim 2$ times
  smaller than suggested by the timing argument, but with large
  dispersion.
  Guided by these
  results, we have selected $12$ halo pairs with total mass in the
  range $1.6$-$3.6 \times 10^{12}\Msun$ for the {\small APOSTLE}
  project (A Project Of Simulating The Local Environment), a suite
  of hydrodynamical resimulations at various numerical resolution levels (reaching up
  to $\sim10^{4}\Msun$ per gas particle) that use the 
  subgrid physics developed for the {\small EAGLE}
  project. These simulations reproduce, by construction, the main
  kinematics of the MW-M31 pair, and produce satellite populations
  whose overall number, luminosities, and kinematics are in good
  agreement with observations of the MW and M31 companions. 
  The {\small APOSTLE} candidate systems thus provide an
  excellent testbed to confront directly many of the predictions of
  the $\Lambda$CDM cosmology with observations of our local Universe.
\end{abstract}



\section{Introduction}
\label{SecIntro}

The Local Group of galaxies (LG), which denotes the association of the
Milky Way (MW) and Andromeda (M31), their satellites, and galaxies in
the surrounding volume out to a distance of $\sim 3$ Mpc, provides a
unique environment for studies of the formation and evolution of
galaxies. Their close vicinity implies that LG galaxies are readily
resolved into individual stars, enabling detailed exploration of the
star formation, enrichment history, structure, dark matter content,
and kinematics of systems spanning a wide range of masses and
morphologies, from the two giant spirals that dominate the Local Group
gravitationally, to the faintest galaxies known.

This level of detail comes at a price, however. The Local Group volume
is too small to be cosmologically representative, and the properties
of its galaxy members may very well have been biased by the peculiar
evolution that led to its particular present-day configuration, in
which the MW and M31, a pair of luminous spirals $\sim 800$ kpc apart,
are approaching each other with a radial velocity of $\sim 120\kms$.
This galaxy pair is surrounded by nearly one hundred galaxies brighter
than $M_V\sim -8$, about half of which cluster tightly around our
Galaxy and M31 \citep[see, e.g.,][for a recent
review]{McConnachie2012}.

The Local Group is also a relatively isolated environment whose
internal dynamics are dictated largely by the MW-M31 pair. Indeed,
outside the satellite systems of MW and M31, there are no galaxies
brighter than $M_B=-18$ (the luminosity of the Large Magellanic Cloud,
hereafter LMC for short) within $3$ Mpc from the MW.  The nearest
galaxies comparable in brightness to the MW or M31 are just beyond $3.5$ Mpc
away (NGC~5128 is at $3.6$ Mpc; M81 and NGC~253 are located $3.7$ Mpc from the MW).

Understanding the biases that this particular environment may induce
on the evolution of LG members is best accomplished through detailed
numerical simulations that take these constraints directly into
account. This has been recognized in a number of recent studies, which
have followed small volumes tailored to resemble, in broad terms, the
Local Group \citep[see,
e.g.,][]{Gottlober2010,Garrison-Kimmel2014}. This typically means
selecting $\sim 3$ Mpc-radius regions where the mass budget is
dominated by a pair of virialized halos separated by the observed
MW-M31 distance and whose masses are chosen to match various
additional constraints \citep[see, e.g.,][]{Forero-Romero2013}.

The mass constraints may include estimates of the individual
virial\footnote{We define the virial mass, $M_{200}$, as
  that enclosed by a sphere of mean density $200$ times the critical
  density of the Universe, $\rho_{\rm crit}=3H^2/8\pi G$. Virial
  quantities are defined at that radius, and are identified by a
  ``200'' subscript.} masses
of both MW and M31, typically based on the kinematics of tracers such
as satellite galaxies, halo stars, or tidal debris \citep[see,
e.g.,][for some recent
studies]{Battaglia2005,Sales2007a,Smith2007,Xue2008,Watkins2010,Deason2012,Boylan-Kolchin2013,Piffl2014,Barber2014}. However,
these estimates are usually accurate only for the mass enclosed within
the region that contains each of the tracers, so that virial mass
estimates are subject to non-negligible, and potentially uncertain,
extrapolation.

Alternatively, the MW and M31 stellar masses may be combined with
``abundance matching'' techniques to derive virial masses \citep[see,
e.g.,][and references therein]{Guo2010,Behroozi2013,Kravtsov2014}. In
this procedure, galaxies of given stellar mass are assigned the virial
mass of dark matter halos of matching number density, computed in a
given cosmological model. Shortcomings of this method include its
reliance on the relative ranking of halo and galaxy mass in a
particular cosmology, as well as the assumption that the MW and M31
are average tracers of the halo mass-galaxy mass relation.

A further alternative is to use the kinematics of LG members to
estimate virial masses. One example is the ``numerical action'' method
developed by \citet{Peebles2001} to reconstruct the peculiar
velocities of nearby galaxies which, when applied to the LG, predicts
a fairly large circular velocity for the MW \citep{Peebles2011}. A
simpler, but nonetheless useful, example is provided by the ``timing
argument'' \citep{Kahn1959}, where the MW-M31 system is approximated
as a pair of isolated point masses that expand radially away after the
Big Bang but decelerate under their own gravity until they turn around
and start approaching. Assuming that the age of the Universe is known,
that the orbit is strictly radial, and that the pair is on first
approach, this argument leads to a robust and unbiased estimate of the
total mass of the system \citep{Li2008}. Difficulties with this
approach include the fact that the tangential velocity of the pair is
neglected \citep[see, e.g.,][]{Gonzalez2014}, together with
uncertainties relating the total mass of the point-mass pair to the
virial masses of the individual systems.

Finally, one may use the kinematics of the outer LG members to
estimate the total mass of the MW-M31 pair, since the higher the mass,
the more strongly LG members should have been decelerated from the
Hubble flow. This procedure is appealing because of its simplicity but
suffers from the uncertain effects of nearby massive structures, as
well as from difficulties in accounting for the directional dependence
of the deceleration and, possibly, for the gravitational torque/pull
of even more distant large-scale structure \citep[see, e.g.,][for some
recent work on the topic]{Penarrubia2014,Sorce2014}.

A review of the literature cited above shows that these methods
produce a range of estimates (spanning a factor of $2$ to $3$) of the
individual masses of the MW and M31 and/or the total mass of the Local
Group \citep[see][for a recent compilation]{Wang2015}. This severely
conditions the selection of candidate Local Group environments that
may be targeted for resimulation, and is a basic source of uncertainty
in the predictive ability of such simulations. Indeed, varying the
mass of the MW halo by a factor of $3$, for example, would likely lead
to variations of the same magnitude in the predicted number of
satellites of such systems \citep{Boylan-Kolchin2010,Wang2012a,Cautun2014},
limiting the insight that may be gained from direct quantitative
comparison between simulations and observations of the Local Group.

With these caveats in mind, this paper describes the selection
procedure, from a simulation of a large cosmological volume, of $12$
viable Local Group environment candidates for resimulation. These $12$
candidate systems form the basis of the {\small EAGLE-APOSTLE}
project, a suite of high-resolution cosmological hydrodynamical
resimulations of the LG environment in the $\Lambda$CDM cosmogony. 
The goal of this paper is to motivate the particular choices made for this
selection whilst critically reviewing the constraints on the total
mass of the Local Group placed by the kinematics of LG members.
Preliminary results from the project (which we shall hereafter refer
to as {\small APOSTLE}, a shorthand for ``A Project Of Simulating 
The Local Environment''), which uses the same code developed for
the {\small EAGLE} simulations \citep{Schaye2015,Crain2015}, have
already been reported in \citet{Sawala2014,Sawala2015} and
\citet{Oman2015}.

\begin{table*}
\caption{The parameters of the cosmological simulations used in this paper.}
\bc
\begin {tabular}{| *{8}{l}  *{3}{c}|}
\hline
Simulation & Cosmology  &   $\Omega_{m}$ &   $\Omega_{\Lambda}$   &
                                                                    $\Omega_{b}$ & $h$    & $\sigma_{8}$ & $n_{s}$  & Cube side  & Particle number & $m_{{\rm p, DM}}$  \\
                    &             &               &                      &             &        &             &          & (Mpc)     &             & ($\Msun$) \\
\hline
MS-I          & WMAP-1     &    0.25       &        0.75          &  0.045      & 0.73   & 0.9        & 1   & 685  & $2160^3$ & $1.2\times10^{9}$ \\
MS-II          & WMAP-1     &    0.25       &        0.75          &  0.045      & 0.73   & 0.9        & 1   & 137  & $2160^3$ & $9.4\times10^{6}$ \\
{\small EAGLE (L100N1504)}        & Planck     &    0.307      &        0.693         &  0.04825     & 0.6777  & 0.8288      & 0.9611 & 100 & $1504^3$ & $9.7\times10^6$\\   
{\small DOVE}    & WMAP-7 &  0.272 &   0.728         &  0.0455     & 0.704  & 0.81       & 0.967 & 100  & $1620^3$ & $8.8\times10^6$  \\
\hline
\end{tabular}
\ec
\label{TabCosmology}
\end{table*}

We begin by using the Millennium Simulations
\citep{Springel2005a,Boylan-Kolchin2009} to select relatively isolated
halo pairs separated by roughly the distance between the MW and M31 and
derive the distribution of total masses of pairs that reproduce,
respectively, the relative radial velocity of the MW-M31 pair, or its
tangential velocity, or the Hubble flow deceleration of distant LG
members. 

Given the disparate preferred masses implied by each of these criteria
when applied individually, we decided to select pairs within a narrow
range of total mass that match loosely the LG kinematics rather than
pairs that match strictly the kinematic criteria but that span the
(very wide) allowed range of masses. This choice allows us to explore
the ``cosmic variance'' of our results given our choice of LG mass,
whilst guiding how such results might be scaled to other possible
choices.  We end by assessing the viability of our candidate selection
by comparing their satellite systems with those of the MW and M31
galaxies.

The plan for this paper is as follows. We begin by assessing in
Sec.~\ref{SecLGMass} the constraints on the LG mass placed by the
kinematics of the MW-M31 pair and other LG members. We describe next,
in Sec.~\ref{SecLGSims}, the choice of {\small APOSTLE} candidates
and the numerical resimulation procedure. Sec.~\ref{SecLGSats}
analyzes the properties of the satellite systems of the main galaxies
of the LG resimulations and compares them with observed LG
properties. We end with a brief summary of our main conclusions in
Sec.~\ref{SecConc}.

\section{The mass of the Local Group}
\label{SecLGMass}

\subsection{Observational Data}

We use below the positions, Galactocentric distances,
line-of-sight-velocities and V-band magnitudes (converted to stellar
masses assuming a mass-to-light ratio of unity in solar units) of
Local Group members as given in the compilation of
\citet{McConnachie2012}.  We also use the relative tangential velocity
of the M31-MW pair derived from M31's proper motion by
\citet{vanderMarel2012}. When needed, we assume an LSR velocity of
$220\kms$ at a distance of $8.5$ kpc from the Galactic center and that
the Sun's peculiar motion relative to the LSR is $U_\odot=11.1$,
$V_\odot=12.24$ and $W_\odot=7.25~\kms$~\citep{Schonrich2010}, to
refer velocities and coordinates to a Galactocentric reference frame.

\begin{figure*}
  \bc \hspace{-0.2cm}
  \resizebox{17cm}{!}{\includegraphics{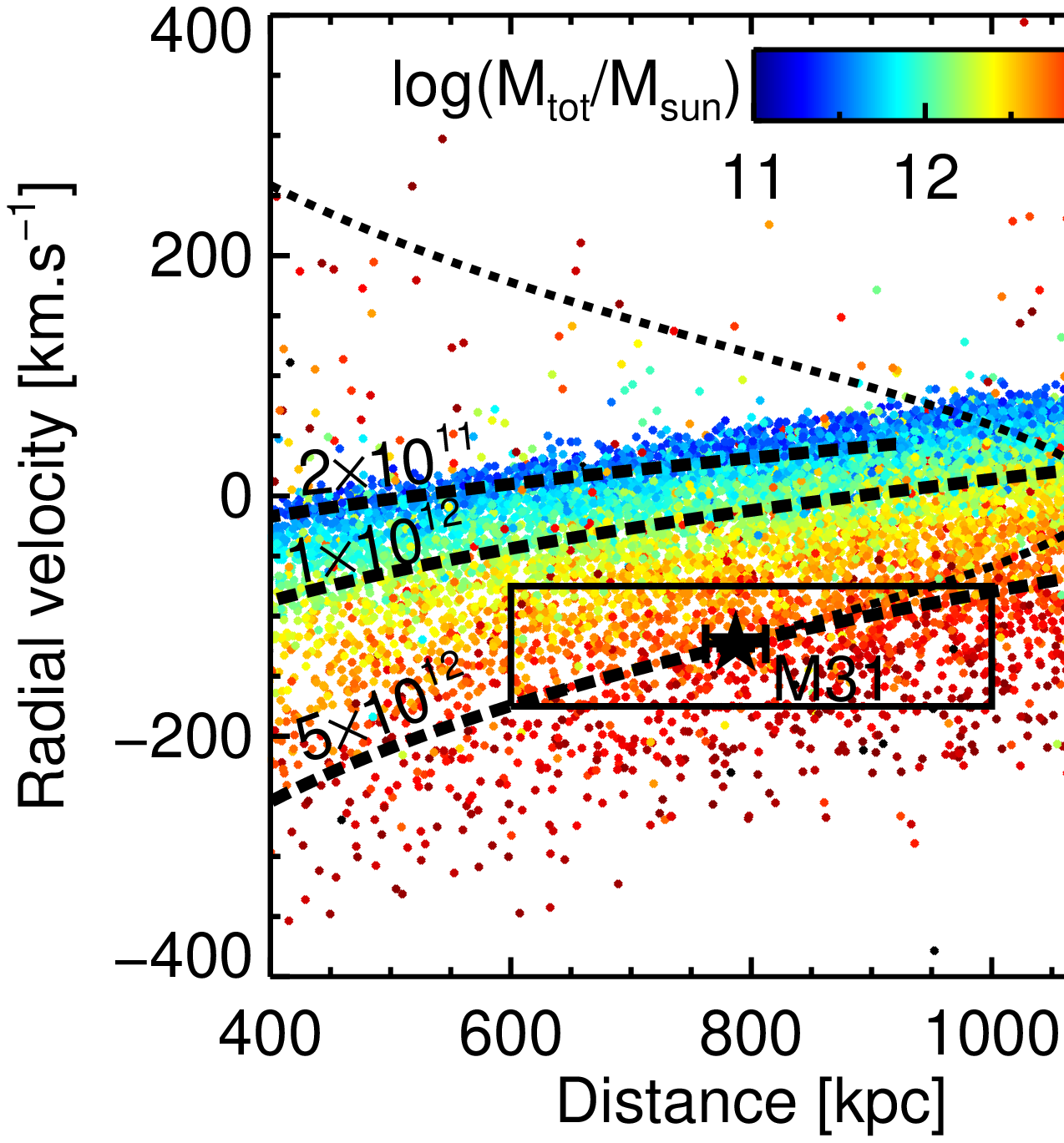}}\\%
  \caption{{\it Left}: The relative radial velocity vs distance of all
    halo pairs selected at highest isolation (``HiIso'') from the Millennium
    Simulation. Colours denote the total mass of the pairs (i.e., the
    sum of the two virial masses), as indicated by the colour bar. The
    starred symbol indicates the position of the MW-M31 pair in this
    plane. The dotted line illustrates the evolution of a point-mass
    pair of total mass $\sim 5 \times 10^{12} \, M_\odot$ in this plane (in physical coordinates) which, 
    according to the timing argument, ends at
    M31's position. Dashed lines indicate the loci, at $z=0$, of pairs
    of given total mass (as labelled) but different initial energies,
    according to the timing argument. The box surrounding the M31
    point indicates the range of distances and velocities used to
    select pairs for further analysis. {\it Right:} Same as left
    panel, but for the relative tangential velocity. Dashed curves in
    this case indicate the mean distance-velocity relation for pairs
    of given total mass, as labelled.}
\label{FigVrVt} \ec
\end{figure*}

According to these data, the MW-M31 pair is $787\pm25$~kpc apart, and
is approaching with a relative radial velocity of
$123\pm4\kms$. In comparison, its tangential velocity is quite low:
only $7\kms$  with 1$\sigma$ confidence region $\leq 22\kms$ . 
We shall assume hereafter that these values are comparable
to the relative velocity of the centers of mass of each member of the pairs selected
from cosmological simulations. In other words, we shall ignore the
possibility that the observed relative motion of the MW-M31 pair may
be affected by the gravitational pull of their massive satellites;
i.e., the Magellanic Clouds (in the case of the MW) and/or M33 (in the
case of M31). This choice is borne out of simplicity; correcting for
the possible displacement caused by these massive satellites requires
detailed assumptions about their orbits and their masses, which are fairly
poorly constrained \citep[see, e.g.,][]{Gomez2015}.

We shall also consider the recession velocity of distant LG members,
measured in the Galactocentric frame. This is also done for
simplicity, since velocities in that frame are more straightforward to
compare with velocities measured in simulations. Other work
\citep[see, e.g.,][]{Garrison-Kimmel2014} has used velocities
expressed in the Local Group-centric frame defined by \citet{Karachentsev1996}. This
transformation aims to take into account the apex of the Galactic
motion relative to the nearby galaxies in order to minimize the
dispersion in the local Hubble flow. This correction, however, is sensitive to
the volume chosen to compute the apex, and difficult to replicate in
simulations.

\begin{figure*}
  \hspace{-0.5cm}
  \resizebox{17cm}{!}{\includegraphics{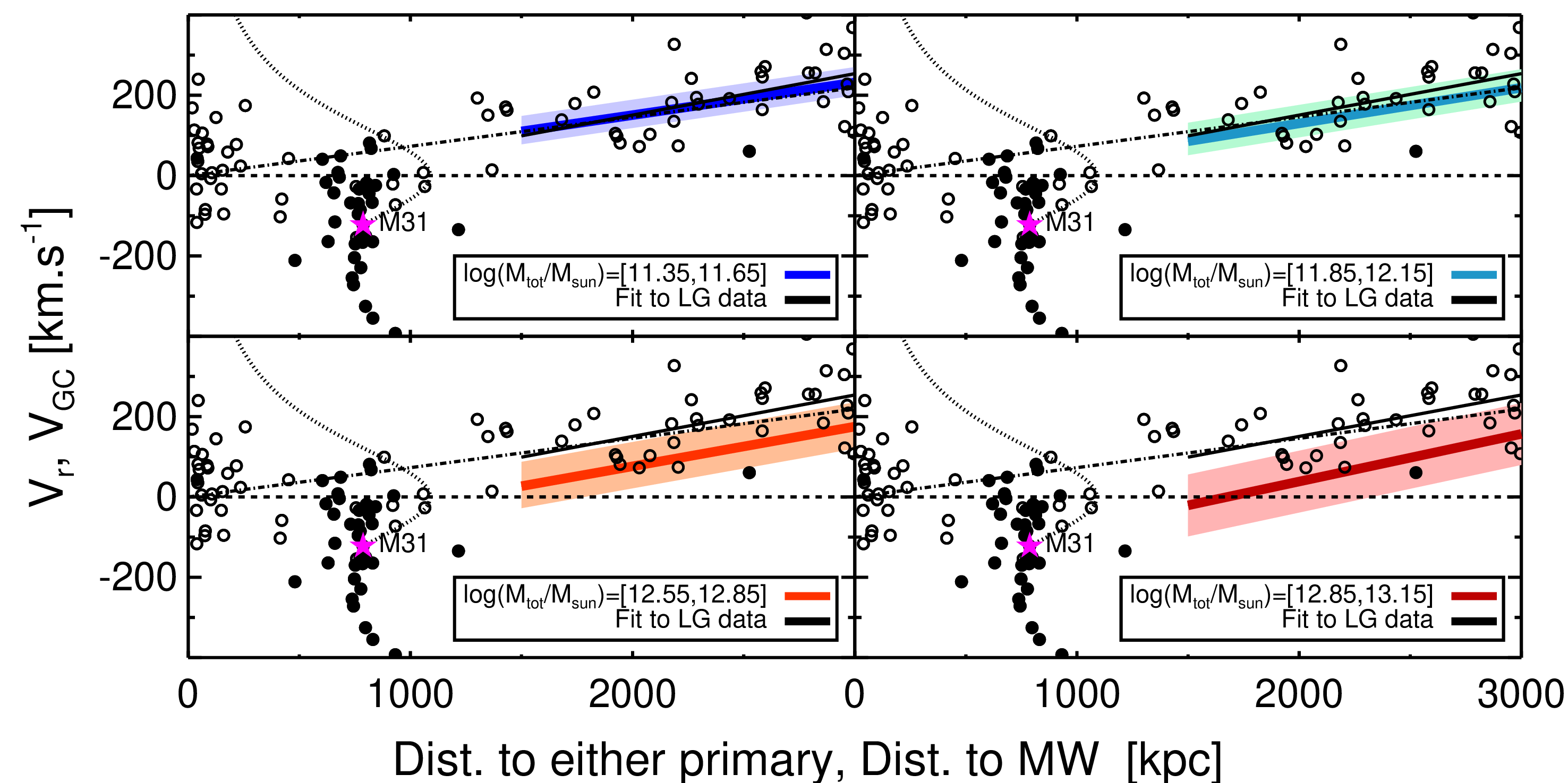}}\\%
  \caption{Radial velocity vs. distance for all Local Group members
    out to a distance of $3$ Mpc from the Galactic center.  Symbols
    repeat in each panel, and correspond to Local Group galaxies in
    the Galactocentric reference frame, taken from the compilations of
    \citet{McConnachie2012} and \citet{Tully2009}. Coloured bands are
    different in each panel and corrrespond to simulations.  Solid
    symbols are used to show LG members that lie, in projection,
    within $30$ degrees from the direction to M31. The dotted line in
    each panel is the timing argument curve for M31, as in
    Fig.~\ref{FigVrVt}. Each panel corresponds to pairs of different
    mass, as given in the legends. The coloured lines and shaded
    regions correspond, in each panel, to the result of binning MedIso
    MS-II halo pairs separated by $600$-$1000$ kpc and least-square
    fitting the recession velocities of the outer members (between
    $1.5$ and $3$ Mpc from either primary). The solid coloured line
    shows the median slope and zero-point of the individual fits. The
    shaded regions show the interquartile range in zero-point
    velocity. Note that the recession speeds of outer members decrease
    with increasing LG mass. The dot-dashed line shows the unperturbed
    Hubble flow, for reference.}
\label{FigHubble}
\end{figure*}

\subsection{Halo pairs from cosmological simulations}
\label{SecSimPairs}

We use the Millennium Simulations, MS-I \citep[][]{Springel2005a} and
MS-II \citep[][]{Boylan-Kolchin2009}, to search for halo pairs with
kinematic properties similar to the MW and M31. The MS-I run evolved
$2160^3$ dark matter particles, each of $1.2 \times 10^9 \Msun$, in a
box $685$ Mpc on a side adopting $\Lambda$CDM cosmological parameters
consistent with the WMAP-1 measurements. The MS-II run evolved a
smaller volume ($137$ Mpc on a side) using the same cosmology and
number of particles as MS-I. Each MS-II particle has a mass of
$9.4 \times 10^6 \Msun$. We list in Table~\ref{TabCosmology} the main
cosmological and numerical parameters of the cosmological simulations
used in our analysis.

At $z=0$, dark matter halos in both simulations were identified using
a friends-of-friends \citep[FoF,][]{Davis1985} algorithm run with a
linking length equal to 0.2 times the mean interparticle
separation. Each FoF halo was then searched iteratively for self-bound
substructures (subhalos) using the {\small SUBFIND} algorithm
\citep{Springel2001b}. Our search for halo pairs include all pairs of
separate FoF halos, as well as single FoF halos with a pair of massive
subhalos satisfying the kinematic and mass conditions we list
below. The latter is an important part of our search algorithm,
since many LG candidates are close enough to be subsumed into a single
FoF halo at $z=0$.

The list of MS-I and MS-II halos retained for analysis include all
pairs separated by $400$~kpc to $1.2$~Mpc whose members have virial
masses exceeding $10^{11}\Msun$ each but whose combined mass does not
exceed $10^{13}\Msun$. These pairs are further required to satisfy a
fiducial isolation criterion, namely that no other halo more massive
than the less massive member of the pair be found within $2.5$~Mpc
from the center of the pair (we refer to this as ``medium isolation''
or ``MedIso'', for short). We have also experimented with
tighter/looser isolation criteria, enforcing the above criterion
within $1$ Mpc (``loosely isolated'' pairs; or ``LoIso'') or $5$~Mpc
(``highly isolated'' pairs; ``HiIso'').  Since the nearest galaxy with
mass comparable to the Milky Way is located at $\sim 3.5$~Mpc, the
fiducial isolation approximates best the situation of the Local Group;
the other two choices allow us to assess the sensitivity of our
results to this particular choice.

\subsection{Radial velocity constraint and the timing argument}
\label{SecTA}

The left panel of Fig.~\ref{FigVrVt} shows the radial velocity vs
separation of all pairs in our MS-I samples, selected using our
maximum isolation criterion. Each point in this panel is coloured by
the total mass of the pairs, defined as the sum of the virial
masses of each member. 

The clear correlation seen between radial velocity and mass, at given
separation, is the main prediction of the ``timing argument''
discussed in Sec.~\ref{SecIntro}. Timing argument predictions are
shown by the dashed lines, which indicate the expected relation for
pairs with total mass as stated in the legend. This panel shows
clearly that low-mass pairs as distant as the MW-M31 pair are still,
on average, expanding away from each other (positive radial velocity),
in agreement with the timing argument prediction (top dashed
curve). Indeed, for a total mass as low as $2\times 10^{11}\Msun$, the
binding energy reaches zero at $r=914$ kpc, $V_r=43\kms$, where the
dashed line ends.

It is also clear that predominantly massive pairs have approach speeds as
large as the MW-M31 pair ($\sim -120\kms$). We illustrate this with
the dotted line, which shows the evolution in the $r$-$V_r$ plane of a
point-mass pair of total mass $5\times 10^{12}\Msun$, selected to
match the MW-M31 pair at the present time. The pair reached
``turnaround'' (i.e., null radial velocity) about $5$ Gyr ago and has
since been approaching from a distance of $1.1$ Mpc (in physical
units) to reach the point labelled ``M31'' by z=0, in the left-panel
of Fig.~\ref{FigVrVt}.

An interesting corollary of this observation is that the present
turnaround radius of the Local Group is expected to be well
beyond $1.1$ Mpc. Assuming, for guidance, that the turnaround radius
grows roughly like $t^{8/9}$ \citep{Bertschinger1985}, this would
imply a turnaround radius today of roughly $\sim 1.7$ Mpc, so that all
LG members just inside that radius should be on first approach. We shall
return to this point when we discuss the kinematics of outer LG
members in Sec.~\ref{SecHubble} below.

\subsection{Tangential velocity constraints}
\label{SecTangVel}

The right-hand panel of Fig.~\ref{FigVrVt} shows a similar exercise to
that described in Sec.~\ref{SecTA}, but using the relative tangential
velocity of the pairs. This is compared with that of the MW-M31 pair,
which is measured to be only $\sim 7\kms$ by \citet{Sohn2012} and is
shown by the starred symbol labelled ``M31''.

This panel shows that, just like the radial velocity, the tangential
velocity also scales, at a given separation, with the total mass of the
pair. In general, higher mass pairs have higher speeds, as gleaned
from the colours of the points and by the three dashed lines, which
indicate the average velocities of pairs with total mass as labelled.

The average relative tangential velocity of a $5\times 10^{12}\Msun$
pair separated by $\sim 800$ kpc is about $\sim 100\kms$, and very few
of such pairs (only $\sim 6\%$) have velocities as low as that of the
MW-M31 pair. The orbit of a typical pair of such mass is thus quite
different from the strictly radial orbit envisioned in timing argument
estimates. Indeed, the low tangential velocity of the MW-M31 pair
clearly favours a much lower mass for the pair than derived from the timing
argument \citep[see][for a similar finding]{Gonzalez2014}.

The kinematics of the MW-M31 pair is thus peculiar compared with that
of halo pairs selected from cosmological simulations: its radial
velocity is best matched with relatively large masses, whereas its
tangential velocity suggests a much lower mass. We shall return to this issue
in Sec.~\ref{SecMassDist}, after considering next the kinematics of
the outer LG members.

\begin{figure}
  \hspace{-0.2cm}
  \resizebox{8cm}{!}{\includegraphics{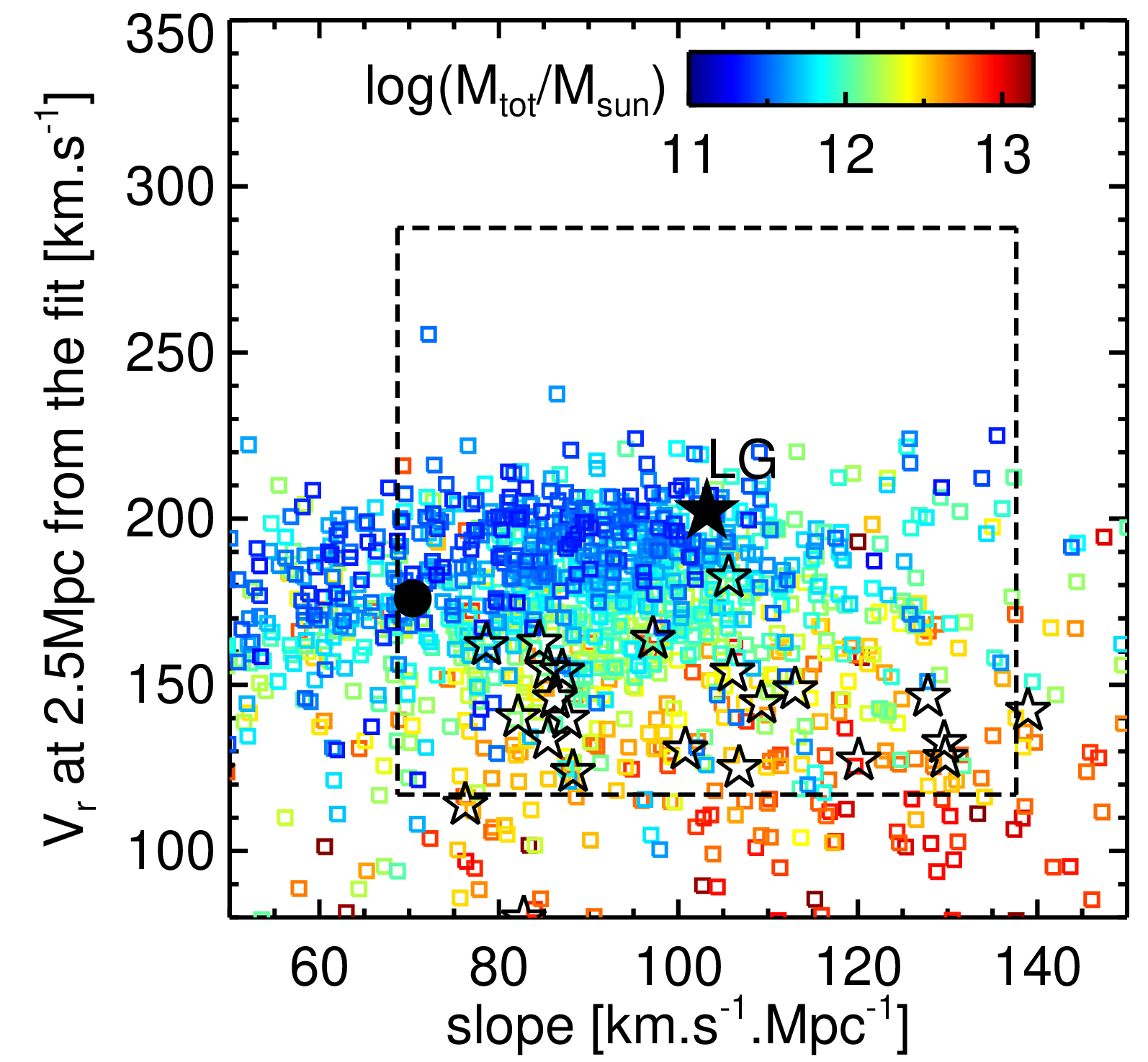}}\\%
  \caption{Parameters of the least-squares fits to the recession speeds
    of outer LG members as a function of distance. 
    The solid starred symbol labelled ``LG''
    indicates the result of fitting the distance vs Galactocentric
    radial velocity of all LG galaxies between $1.5$ and $3$ Mpc from
    the MW. The Hubble flow is shown by the solid circle. Each
    coloured point corresponds to a halo pair selected from the MS-II simulation
    assuming medium isolation, and uses all halos in the $1.5$-$3$ Mpc
    range resolved in MS-II with more than $100$ particles (i.e.,
    masses greater than $1\times 10^{9}\Msun$).  Velocities and
    distances are measured from either primary, and coloured according
    to the total mass of the pair. Note that recession velocities
    decrease steadily as the total mass of the pair increases. The
    square surrounding the ``LG'' point indicates the error
    in the slope and zero-point, computed directly from the fit to the
    $33$ LG members with distances between $1.5$ and $3$ Mpc. 
    Open starred symbols correspond to the $12$ pairs selected for the
    {\small APOSTLE} project (see Sec.~\ref{SecCandidates}).}
\label{FigHubbleFits}
\end{figure}

\subsection{Hubble flow deceleration}
\label{SecHubble}

The kinematics of the more distant LG members is also sensitive to the
total mass of the MW-M31 pair. In particular, we expect that the
larger the mass, the more the recession velocities of those members
would be decelerated from the Hubble flow.  We explore this in
Fig.~\ref{FigHubble}, where the symbols in each panel show the
Galactocentric radial velocity of all galaxies in the
\citet{McConnachie2012} catalogue and in the Extragalactic Distance
Database of \citet{Tully2009} found within $3$ Mpc of the MW. These
data illustrate two interesting points. One is that all galaxies
beyond $\sim 1.3$ Mpc from the MW are receding; and the second is that
the mean velocity of all receding galaxies is only slightly below the
Hubble flow, $V_r=H_0 r$, indicated by the dot-dashed line for
$H_0=70.4\kmsMpc$.

The first result is intriguing, given our argument in Sec.~\ref{SecTA}
that the turnaround radius of the LG should be around $\sim 1.7$ Mpc
at the present time. The second point is also interesting, since it
suggests that the local Hubble flow around the MW (beyond $\sim
1.3$ Mpc) has been relatively undisturbed. The first point suggests
that the M31 motion is somewhat peculiar relative to that of the rest
of the LG members; the second that the total mass of the MW-M31 pair
cannot be too large, for otherwise the recession velocities would have been
decelerated more significantly.

We examine this in more detail in the four panels of
Fig.~\ref{FigHubble}, where the coloured lines show the mean recession
speed as a function of distance for halos and subhalos surrounding
candidate MS-II\footnote{We use here only MS-II pairs; the numerical
  resolution of the MS-I simulation is too coarse to identify enough
  halos and subhalos around the selected pairs to accurately measure the
  velocities of outer LG members.} pairs of different mass, binned as listed in the
legend. Recession speeds are measured from each of the primaries for
systems in the distance range $1.5$-$3$ Mpc and are least-square
fitted independently to derive a slope and velocity zero-point for
each pair. The coloured lines in Fig.~\ref{FigHubble} are drawn using
the median slope and zero-point for all pairs in each mass bin. The
shaded areas indicate the interquartile range in the zero-point
velocity. This figure shows clearly that the more massive the pair,
the lower, on average, the recession velocities of surrounding
systems.

This is also clear from Fig.~\ref{FigHubbleFits}, where we plot the
individual values of the slopes and zero-point velocities (measured at
$r=2.5$ Mpc), compared with the values obtained from a least-squares
fit to the Local Group data. The zero-point of the fit is most
sensitive to the total mass of the pair; as a result, the relatively
large recession velocity of the outer LG members clearly favours a low total mass
for the MW-M31 pair.

\begin{figure}
  \bc \hspace{-0.2cm}
  \resizebox{8.5cm}{!}{\includegraphics{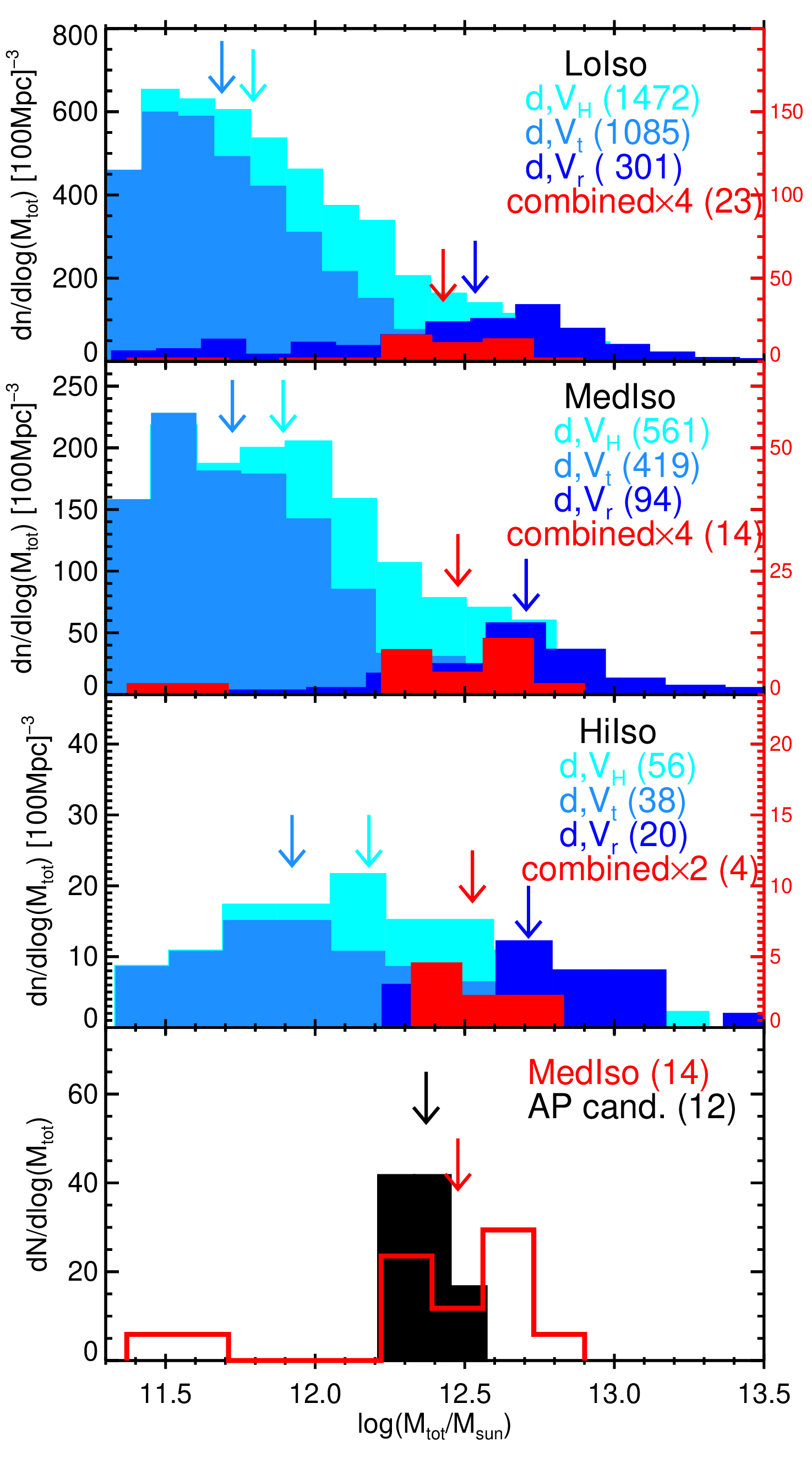}}\\%
  \caption{Total mass distributions of all pairs in the Millennium-II
    Simulation (MS-II) with separation and relative radial velocity
    similar to the MW-M31 pair, selected assuming different isolation
    criteria. The dark blue histogram shows pairs that satisfy the
    distance-radial velocity criterion shown by the box in the left
    panel of Fig.~\ref{FigVrVt}. The lighter blue histogram shows
    those pairs satisfying the constraint shown by the box in the
    distance-tangential velocity panel of Fig.~\ref{FigVrVt}. The
    lightest blue histogram identifies pairs satisfying the ``Hubble
    flow'' criterion shown by the box in Fig.~\ref{FigHubbleFits}. The
    red histogram corresponds to the intersection of all three
    criteria; their scale has been increased to make them more easily
    visible. The axis on the right shows the scale of the red histogram. 
    From top to bottom, the first three panels show results for
    different levels of isolation, as labelled. The numbers in
    brackets indicate the number of pairs in each histogram. Arrows
    indicate the median of each distribution. The bottom panel shows
    in red the results for medium isolation (empty histogram),
    and compares them with the mass distribution of the $12$ candidate
    pairs selected from the {\small DOVE} simulation for the {\small
      APOSTLE} simulation project (in solid black).}
\label{FigMassHist} \ec
\end{figure}

\subsection{Mass distributions}
\label{SecMassDist}

The histograms in Fig.~\ref{FigMassHist} summarize the results of the
previous three subsections. Each of the top three panels shows the
distribution of the total masses of pairs with separations in the
range ($600,1000$) kpc selected to satisfy the following constraints:
(i) relative radial velocity in the range ($-175,-75$)~$\kms$ (dark
blue histograms; see selection box in the left panel of
Fig.~\ref{FigVrVt}); (ii) tangential velocity in the range ($0$,$50$)
$\kms$ (light-blue histogram; see box in the right panel of
Fig.~\ref{FigVrVt}); and (iii) Hubble velocity fits in the observed
range (lightest blue histograms; see box in
Fig.~\ref{FigHubbleFits}). The histograms in red correspond to the few
systems that satisfy all three criteria at once. (The scale of the red
histograms has been increased in order to make them visible; see
legends.) In these three panels the isolation criteria for selecting
pairs tightens from top to bottom, and, as a consequence, the total
number of selected pairs decreases.  As discussed before, the radial
velocity constraint favours high-mass pairs, whereas the tangential
velocity constraint favours low-mass ones. The Hubble velocity
constraint gives results intermediate between the other two.

  Fig.~\ref{FigMassHist} shows that the main effect of relaxing the
  isolation criteria is to enable relatively low-mass pairs to pass
  the kinematic selection. The total mass of ``HiIso'' pairs matching
  the radial velocity constraint, for example, clusters tightly about
  the timing argument prediction ($\sim 5\times 10^{12}\, M_\odot$;
  see dark blue histogram in the ``HiIso'' panel). However, a long
  tail of pairs with much lower masses appears in the other panels,
  indicating that neighbouring structures can have a non-negligible
  effect on the kinematics of a pair. In particular, ``infall'' onto a
  massive structure may accelerate low-mass pairs to much
  higher relative velocities than they would be able to reach in isolation.

  The magnitude of the effect is expected to scale with the mass and
  distance to the most massive neighbouring system, and it seems
  legitimate to question whether an object like the Virgo cluster
  might have a discernible effect on the kinematics of the MW-M31
  pair. We have checked this explicitly {\bf by identifying the
    subsample of pairs in Fig.~\ref{FigMassHist} that have a halo of
    mass $\gsim 5\times 10^{13}\, M_\odot$ in the distance
    range\footnote{These numbers are chosen to match the Virgo
      cluster, a $\sim 4 \times 10^{14}\ M_\odot$ system situated $17$
      Mpc from the MW \citep{McLaughlin1999,Tonry2001}.} $15$-$20$
      Mpc. We find no statistically significant difference between
    the pair mass distribution of this subsample and that of the
    MedIso and LoIso samples shown in Fig.~\ref{FigMassHist}. (This
    subsample is too small to draw a statistically sound conclusion
    for the HiIso case.)}  This suggests that only nearby massive
    systems can skew the pair mass distribution, and that a cluster as
    distant as Virgo is unlikely to play a substantial role in the
    kinematics of the M31-MW pair.


A further feature highlighted by Fig.~\ref{FigMassHist} is that very
few pairs satisfy all three constraints simultaneously. For the case
of medium isolation (second panel from the top in
Fig.~\ref{FigMassHist}) only $14$ pairs satisfy the three criteria
simultaneously in the $(137$ Mpc$)^3$ volume of the MS-II simulation.
The bottom panel of Fig.~\ref{FigMassHist} shows the mass distribution
of pairs that satisfy all three criteria in the case of ``medium
isolation'', the closest to our observed LG configuration (empty red
histogram).  The mass distribution of pairs that satisfy the three
criteria simultaneously peaks at $\sim 2\times 10^{12}\Msun$, a value
significantly lower than the mass suggested by the timing
argument. The distribution is quite broad, with an rms of $0.4$ dex
and a full range of values that stretches more than a decade; from
$2.3\times 10^{11}\Msun$ to $6.1\times 10^{12}\Msun$ in the case of
medium isolation.

This finding is one of the main reasons guiding our choice of mass for
the halo pairs that we select for resimulation in the {\small APOSTLE} 
project; their distribution is shown by the solid black
histogram in the bottom panel of Fig.~\ref{FigMassHist}. We describe
the selection procedure of these $12$ pairs in detail next.

\begin{figure}
  \bc \hspace{-0.2cm}
  \resizebox{8.cm}{!}{\includegraphics{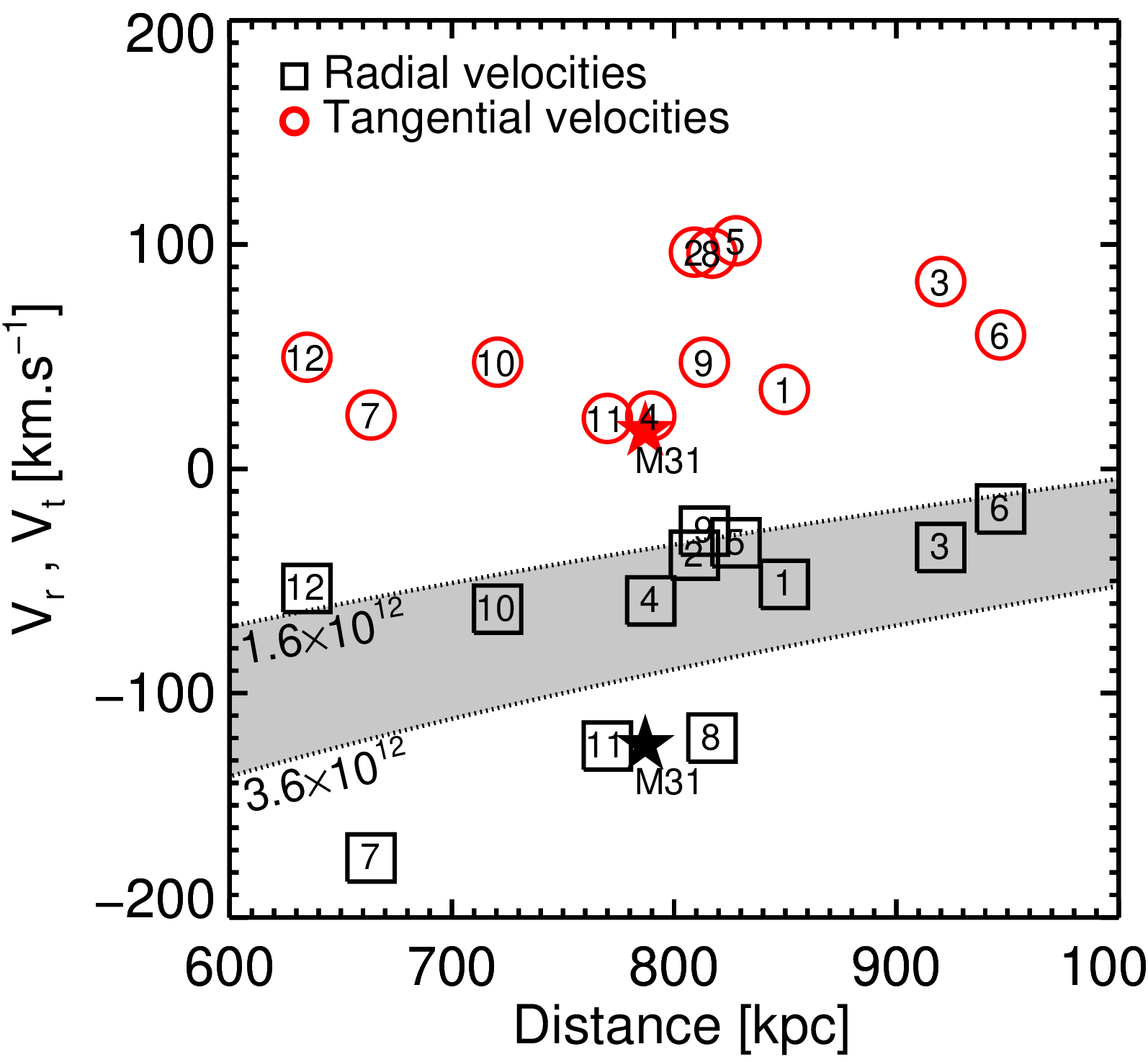}}\\%
  \caption{Relative radial (squares) and tangential (circles)
    velocities vs separation for all $12$ candidates selected for
    resimulation in the {\small APOSTLE} project. The
    corresponding values for the MW-M31 pair are shown by the starred
    symbols. The shaded area indicate the region where we would expect
    the radial velocities of all $12$ candidates to lie if the
    predictions of the timing argument held.}
\label{FigVrVtSims} \ec
\end{figure}

\section{The {\small APOSTLE} simulations}  
\label{SecLGSims}

The {\small APOSTLE} project consists of a suite of
high-resolution cosmological hydrodynamical simulations of $12$ Local
Group-like environments selected from large cosmological volumes of a
$\Lambda$CDM universe. Each of these volumes has been simulated at
three different resolutions with the same code as was used for the
{\small EAGLE} simulation project \citep{Schaye2015,Crain2015}. 

Results from a subset of these simulations have already been presented
in \citet{Sawala2014,Sawala2015} and \citet{Oman2015}. Here we
describe the selection procedure for the resimulation volumes together
with the physical processes included in the simulations. We assess the
viability of these candidates by analyzing, in Sec.~\ref{SecLGSats},
their satellite populations and comparing them with observations of
the satellites of the MW and M31 galaxies.

\subsection{The code}
\label{SecCode}

The {\small APOSTLE} simulations were run using a highly modified
version of the N-body/SPH code, P-Gadget3 \citep{Springel2005b}. The
code includes sub-grid prescriptions for star formation, feedback from
evolving stars and supernovae, metal enrichment, cosmic reionization,
and the formation and energy output from supermassive black
holes/active galactic nuclei (AGN). The details of the code and the
subgrid physics model are described in detail in \citet{Schaye2015}
and \citet{Crain2015} .  When used to evolve a cosmologically
significant volume, this code gives a galaxy stellar mass function in
good agreement with observations in the range of galaxy stellar masses
between $\sim 10^{8}$ and $\sim 10^{11}\Msun$\citep{Schaye2015}. The
{\small APOSTLE} runs use the same parameter choices as the
``reference'' model in \citet{Schaye2015}. As discussed below, we find
good numerical convergence in the galaxy properties we investigate,
without further calibration.

\begin{table*}
  \caption{The parameters of the {\small APOSTLE} resimulations. The
    first two columns list labels identifying each run. The following
    columns list the virial masses of each of the primaries at $z=0$; their
    relative separation, radial velocity, and tangential velocity 
    in the {\small DOVE} simulation, as
    well as the initial baryonic mass per particle in the hydrodynamical
    runs. The dark matter particle mass is $m_{\rm DM}=(1/f_{\rm
      bar}-1) \, m_{\rm gas}$, 
    where $f_{\rm bar}$ is the universal baryon
    fraction. (Dark matter-only runs have a particle mass equal to the
    sum of $m_{\rm gas}+m_{\rm DM}$.) The last column lists the value of the
    Plummer-equivalent gravitational softening, which is comoving at
    early times, but fixed at the listed value after $z=3$.}
  \bc
\begin {tabular}{| *{4}{l}  *{1}{c} *{2}{r} *{2}{c}|}
\hline
Name & Run & $M_{200}^{[1]}$ & $M_{200}^{[2]}$        &separation & $V_{\rm r}$ &
                                                                     $V_{\rm t}$
  & $m_{\rm gas}$   & $\epsilon_{\rm {max}}$  \\
     & (resolution)  & [$10^{12}(\Msun)$]& [$10^{12}(\Msun)]$ &[kpc] & [$\kms$]& [$\kms$]&[$10^{4} (\Msun)$] & [pc]    \\  
\hline
AP-1  &  L1/L2/L3   & $1.66$  &   $1.10$  &  $850$   &   $-51$   & $35$  & $0.99$/$12.0$/$147$ &  $134/307/711$ \\
AP-2  &  L2/L3   & $0.85$ & $0.83$   &  $809$   &   $-39$   & $97$  & $12.5$/$147$ &  307/711    \\
AP-3  &  L2/L3   & $1.52$  &   $1.22$  &  $920$   &   $-35$   & $84$  & $12.5$/$147$ &  307/711     \\
AP-4  &  L1/L2/L3   & $1.38$  &   $1.35$  &  $790$   &   $-59$   & $24$  & $0.49$/$12.2$/$147$ & 134/307/711  \\
AP-5  &  L2/L3   & $0.93$ &   $0.87$ &  $828$   &   $-33$   & $101$ & $12.5$/$147$ &  307/711 \\
AP-6  &  L2/L3   & $2.36$  &   $1.21$  &  $950$   &   $-18$   & $60$  & $12.7$/$137$ &  307/711 \\
AP-7  &  L2/L3   & $1.88$  &   $1.09$  &  $664$   &   $-174$  & $24$  & $11.3$/$134$ &  307/711 \\
AP-8  &  L2/L3   & $1.72$  &  $0.65$  &  $817$   &   $-120$  & $96$  & $11.0$/$137$ &  307/711 \\
AP-9  &  L2/L3   & $0.96$ &  $0.68$  &  $814$   &   $-28$   & $48$  & $10.9$/$138$ &  307/711 \\
AP-10 &  L2/L3   & $1.46$  &   $0.87$ &  $721$   &   $-63$   & $48$  & $11.0$/$146$ &  307/711 \\
AP-11 &  L2/L3   & $0.99$ &   $0.80$ &  $770$   &   $-124$  & $22$  & $11.1$/$153$ &  307/711 \\
AP-12 &  L2/L3   & $1.11$  &  $0.58$  &  $635$   &   $-53$   & $50$  & $10.9$/$138$ &  307/711 \\
\hline
\end{tabular}
\ec
\label{TabLGSims}
\end{table*}

\subsection{Candidate selection}
\label{SecCandidates}

The pairs selected for resimulation in the {\small APOSTLE}
project were drawn from the {\small DOVE}\footnote{{\small DOVE} 
simulation is also known as {\small COLOR}.} cosmological N-body
simulation described by \citet{Jenkins2013}. {\small DOVE} evolved a
periodic box $100$ Mpc on a side assuming cosmological parameters
consistent with the WMAP-7 estimates and summarized in
Table~\ref{TabCosmology}.  {\small DOVE} has $134$ times better mass
resolution than MS-I, with a particle mass of $8.8\times10^{6}\Msun$
(comparable to MS-II).  Halo pairs were identified in {\small DOVE}
using the procedure described in Sec.~\ref{SecSimPairs}.

Guided by the discussion in Sec.~\ref{SecLGMass}, we chose $12$
different pairs from the MedIso sample that satisfied,
at $z=0$, the following conditions:

\begin{itemize}

\item{separation between $600$ and $1000$ kpc,}

\item{relative radial velocity, $V_{\rm r}$, in the range $(-250,0) \kms$,}

\item{relative tangential velocity, $V_{\rm t}$, less than $100 \kms$ ,}

\item{recession velocities of outer LG members in the range defined by
    the box in Fig.~\ref{FigHubbleFits}}

\item{total pair mass (i.e., the sum of the virial masses of the two
    primary halos) in the range $\log(M_{\rm
      tot}/\Msun)=[12.2,12.6]$.}

\end{itemize}

The relative velocities and separations of the 12 pairs are shown in
Fig.~\ref{FigVrVtSims}, where each pair is labelled with a number from
$1$ to $12$ for future reference. The main properties of these pairs
are listed in Table~\ref{TabLGSims} and the histogram of their total
masses is shown (in black) in the bottom panel of
Fig.~\ref{FigMassHist}.  The parameters of fits to the recession
velocities of the outer members are shown by the open starred symbols
in Fig.~\ref{FigHubbleFits}.


Note that the $12$ pairs chosen span a relatively small range of masses
compared to what is allowed according to the kinematic constraints
described in the previous section. The
lowest mass of the pairs is $1.6\times10^{12} \Msun$ and the 
most massive pair is $3.6\times 10^{12} \Msun$,
with an average mass of $2.3\times10^{12} \Msun$ and 
an rms of only $30\%$. The small mass
range of the pairs was chosen in order to explore the
cosmic variance at fixed mass. Given the large range of allowed masses 
it would have been impossible to cover the whole range with
high resolution resimulations.

It is interesting to compare the masses of the pairs with those
estimated from the timing argument applied to each pair. This is shown by the shaded region
of Fig.~\ref{FigVrVtSims}, which brackets the region in
distance-radial velocity plane that should be occupied by the $12$
candidates we selected, given their total masses. The timing argument
does reasonably well in $8$ out of the $12$ cases, but four pairs fall
outside the predicted region.  Three pairs, in particular, have
approach velocities as high or higher than observed for the MW-M31
pair (pairs 7, 8, and 11), demonstrating that our mass choice does not
preclude relatively high approach velocities in some cases.

\subsection{Resimulation runs}

Volumes around the $12$ selected pairs were carefully configured in
the initial conditions of the {\small DOVE} simulation so that the
inner $2$-$3$ Mpc spherical regions centred on the mid-point of the
pairs contain no ``boundary'' particles at $z=0$ \citep[for details of
this ``zoom-in'' technique, see][]{Power2003,Jenkins2013}. For each
candidate pair, initial conditions were constructed at three levels of
resolution using second-order Lagrangian perturbation theory
\citep{Jenkins2010}. (Further details are provided in the Appendix A.)
Resolution levels are labelled L1, L2, and L3 (high, medium, and low
resolution), where L3 is equivalent to {\small DOVE}, L2 is a factor
of $\sim 10$ improvement in mass resolution over L3, and L1 is another
factor of $\sim 10$ improvement over L2. The method used to make the
zoom initial conditions has been described in \citet{Jenkins2013}.
Particles masses, softening lengths, and other numerical parameters
for each resolution level are summarized in Table~\ref{TabLGSims}.


Each run was performed twice, with one run neglecting the baryonic
component (i.e., assigning all the matter to the dark matter
component, hereafter referred to as dark-matter-only or ``DMO'' runs)
and the other using the {\small EAGLE} code described in
Sec.~\ref{SecCode}.  Note that the cosmology assumed for the {\small
  APOSTLE} runs is the same as that of {\small DOVE} (see
Table~\ref{TabCosmology}), which differs slightly from that of {\small
  EAGLE}.  These differences, however, are very small, and are
expected to have a negligible effect on our results. At the time of
writing, only two volumes had been completed at the highest (L1)
resolution; AP-1 and AP-4.

%

\section{Local Group satellites}
\label{SecLGSats}

In this section we explore the properties of the satellite population
that surrounds each of the two primary galaxies of the {\small APOSTLE} 
resimulations to assess whether our choice of total mass and
subgrid physics gives results consistent with the broad properties of
the satellite systems of the MW and M31. We examine consistency with
properties that may be considered problematic, given that our choice
for the total mass is significantly lower than indicated by the timing
argument. In particular, we focus on (i) whether the total number of
satellites brighter than $M_V=-8$ and $M_V=-9.5$ (or, equivalently,
stellar mass\footnote{Stellar masses are measured within a radius,
  $r_{\rm gal}$, equal to $15\%$ of the virial radius, $r_{200}$, of
  the surrounding halo. For satellites we estimate $r_{\rm gal}$ from
  its peak circular velocity, $V_{\rm max}$, and the relation between
  $V_{\rm max}$ and $r_{200}$ for isolated galaxies. }, $M_{\rm gal}$,
greater than $1.4\times 10^5$ or $5.6 \times 10^5\Msun$,
respectively), is comparable to the observed numbers around the MW and
M31; (ii) whether satellites as massive/bright as the Large Magellanic
Cloud or M33 are present; (iii) whether the velocity dispersion of the
satellites is consistent with observations; and, finally, (iv) whether
the observed satellites within $300$ kpc of either primary (i.e., the
MW or M31), are gravitationally bound given our choice of total mass
for the LG candidates.

 \begin{figure}
 \bc \hspace{-0.2cm}
  \resizebox{8.cm}{!}{\includegraphics{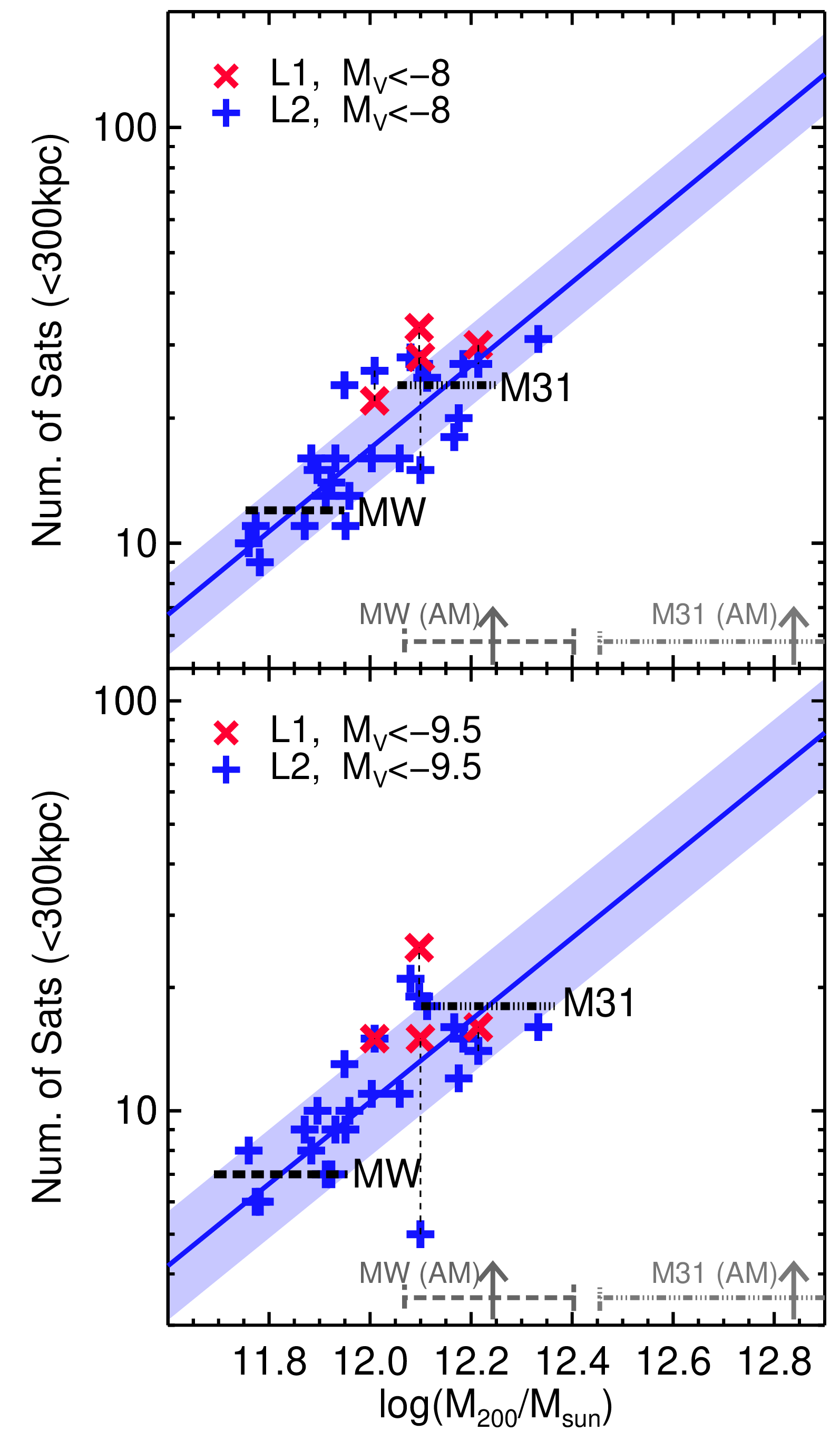}}\\%
  \caption{{\it Top}: Number of satellites with stellar mass greater
    than $1.4\times 10^5 \Msun$ (or brighter than $M_V=-8$) found
    within $300$ kpc of each of the primary galaxies in our {\small
      APOSTLE} runs plotted versus the primary's virial mass. We
    show results for medium resolution (L2, blue plus symbols) and
    high resolution (L1, red crosses). The thin dashed lines connect
    the L1 halos to their counterparts in L2.  The solid, coloured
    lines indicate the best fit to L2 data with unit slope; the
    shaded areas indicate the rms range about this fit. Small
    horizontal lines indicate the observed numbers for the MW and
    M31. Arrows indicate virial masses estimated for the MW and M31
    from abundance matching of \citet{Guo2010}. The ``error bar''
    around each arrow spans different predictions by
    \citet{Behroozi2013} and \citet{Kravtsov2014}.  {\it Bottom}: Same
    as the top panel but for satellites with stellar mass greater than
    $5.6\times 10^5 \Msun$ (or brighter than $M_V=-9.5$). }
\label{FigNSats}\ec
\end{figure}

 \begin{figure}
  \bc \hspace{-0.2cm}
  \resizebox{8.cm}{!}{\includegraphics{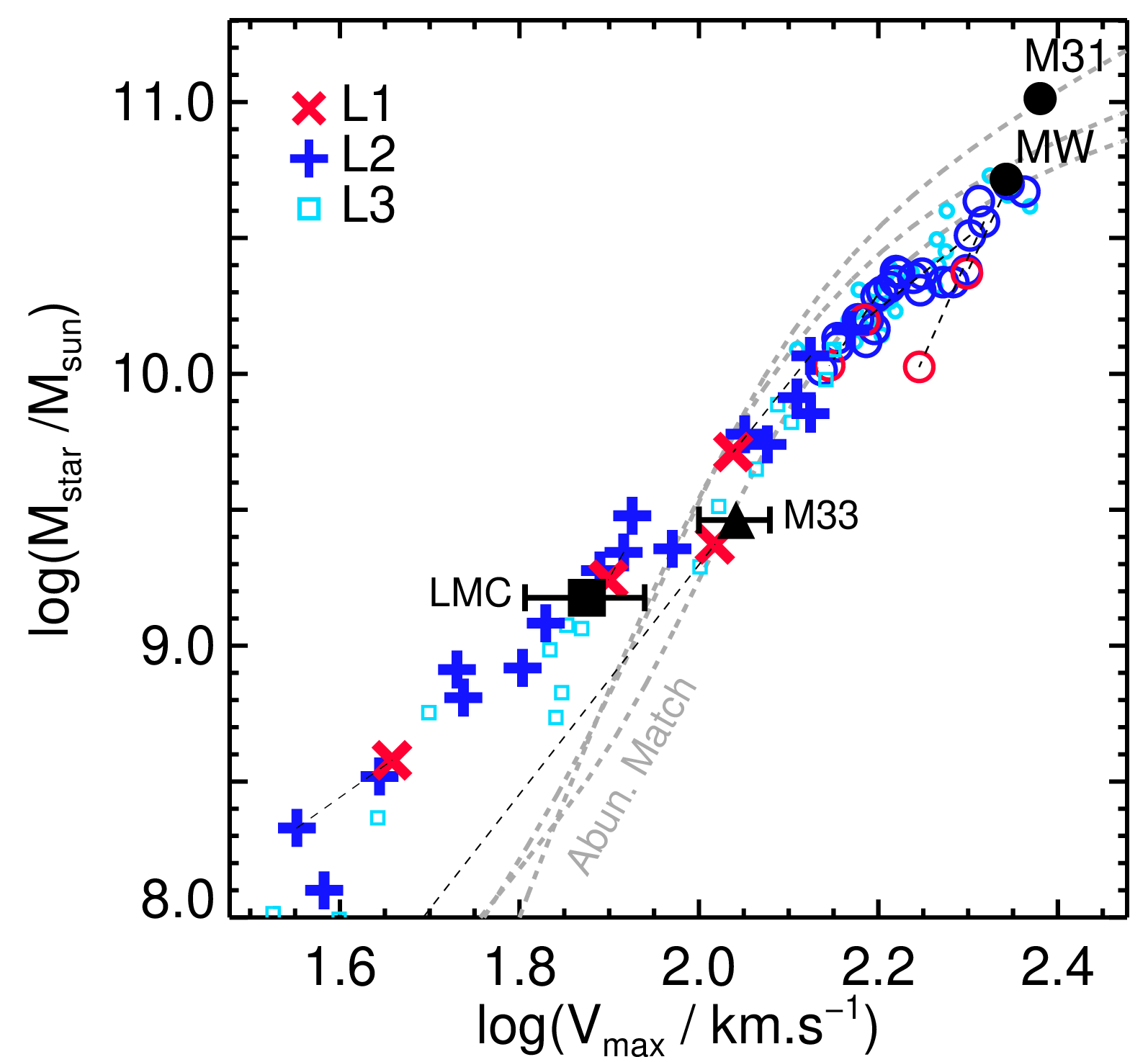}}\\%
  \caption{Stellar mass of the primary galaxies (circles) and their
    brightest satellites (crosses) for our $12$ {\small APOSTLE}
    simulations, as a function of their maximum circular velocity. The large
    black solid symbols denote observational estimates for the two
    brightest members of the Local Group and their brightest
    satellites: MW, M31, LMC, and M33, respectively. Grey curves
    indicate ``abundance matching'' predictions from
    \citet{Guo2010,Behroozi2013,Kravtsov2014}. Note that satellites as
    bright and massive as the LMC and M33 are not uncommon in our
    simulations. Note also that our primaries tend to be undermassive
    relative to observations and to predictions from abundance
    matching. The reverse applies at lower circular velocities. Black
    dashed lines connect matching systems at different resolution
    levels. See text for further discussion. }
\label{FigBrightSats}\ec
\end{figure}

\subsection{Satellite masses/luminosities}
\label{SecSatLums}

\subsubsection{Number of bright satellites}
\label{SecNsat}

The total number of satellites above a certain mass is expected to be
a sensitive function of the virial masses of the primary halos
\citep[see, e.g.,][]{Boylan-Kolchin2010,Wang2012a}, so we begin by
considering the number of satellites whose stellar masses exceed
$1.4\times 10^5 \Msun$ or, equivalently\footnote{We assume a constant
  mass-to-light ratio of $1$ in solar units for converting stellar
  masses to $V$-band luminosities. This is done only for simplicity,
  since our primary aim is to assess consistency rather than to
  provide quantitative predictions. Later work will use
  spectrophotometric models and internal extinction as laid out in
  \citet{Trayford2015}.}, that are brighter than $M_V=-8$, as a
function of the virial mass ,$M_{200}$, of each host.  This brightness
limit corresponds roughly to the faintest of the ``classical'' dwarf
spheroidal companions of the MW, such as Draco or Ursa Minor. We
choose this limit because there is widespread consensus that surveys
of the surroundings of the MW and M31 are complete down to that limit
\citep[see, e.g.,][]{Whiting2007,McConnachie2009,McConnachie2012}.

We compare in the top panel of Fig.~\ref{FigNSats} the number of
simulated satellites within $300$ kpc of each primary galaxy to the
numbers observed around the MW and M31 (shown as horizontal line
segments). Satellite numbers correlate strongly with virial mass, as
expected, and it is encouraging that the observed number of satellites
of the MW and M31 (shown by short horizontal line segments in
Fig.~\ref{FigNSats}) are well within the range spanned by our
simulations. Since a lower-mass limit of
$M_{\rm gal}=1.4\times 10^5\Msun$ corresponds to just a few
particles at L2 resolution, we repeat the exercise in the
bottom panel of Fig.~\ref{FigNSats}, but for a higher mass limit;
i.e., $M_{\rm gal}>5.6\times 10^5\Msun$ ($M_V<-9.5$). The results 
in either panel are reassuringly similar.

The impact of numerical resolution may be seen by comparing the
results for the medium resolution (L2; blue ``$+$'' symbols) with
those obtained for the L1 (high-resolution) runs in
Fig.~\ref{FigNSats} (red crosses). On average, the number of
satellites increases by only $\sim 10\%$ when increasing the mass
resolution by a factor of $10$, indicating rather good
convergence. One of the halos in L1, however, hosts almost twice as
many satellites as its counterpart in L2. The reason is that a
relatively large group of satellites has just crossed inside the $300$
kpc boundary of the halo in L1, but is still outside $300$ kpc in
L2. (We do not consider the low-resolution L3 runs in this plot
because at that resolution the mass per particle is
$1.5\times 10^{6}\Msun$ and satellites fainter
  than $M_V\sim-13$ are not resolved, see Table~\ref{TabLGSims}.)

We also note that, had we chosen larger masses for our LG primaries, 
consistent with the timing argument and abundance matching,
our simulations would have likely formed a much larger number of bright
satellites than observed. For example, for a virial mass of
$\sim 2 \times 10^{12}\Msun$ (the value estimated for the MW 
by abundance-matching analyses; see the upward arrow in
Fig.~\ref{FigNSats}) our simulations give, on average, $25$
satellites brighter than $M_V=-8$ and $15$ satellites brighter than
$M_V=-9.5$. These are well in excess of the $12$ and $8$ satellites,
respectively, found in the halo of the MW. The same conclusion
applies to M31, where abundance matching suggests a halo mass of order
$7\times 10^{12}\Msun$. It is clear from Fig.~\ref{FigNSats} that
our simulations would have produced a number of satellites well in
excess of that observed around M31 for a halo as massive as that.

The correlation between host virial mass and
satellite number is, of course, sensitive to our choice of galaxy
formation model, and it would be possible, in principle, to reduce the
number of bright satellites by reducing the overall galaxy formation
efficiency in low-mass halos. It is interesting that the
same galaxy formation model able to reproduce the galaxy stellar mass
function in large cosmological volumes \citep{Schaye2015} can
reproduce, without further tuning, the number of satellites in the
{\small APOSTLE} resimulations.


\begin{figure}
  \bc \hspace{-0.2cm}
  \resizebox{8.cm}{!}{\includegraphics{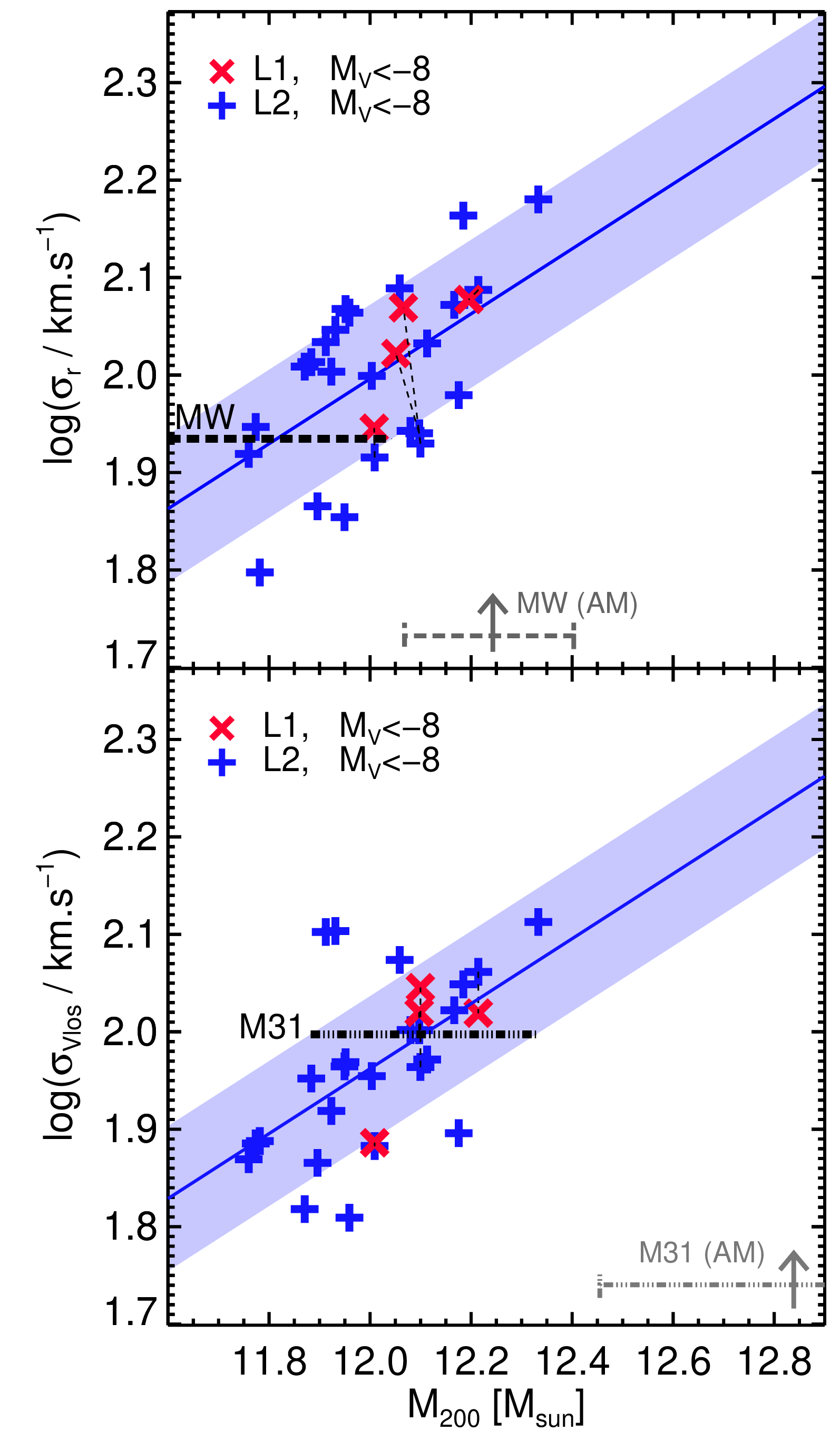}}\\%
  \caption{{\it Top}: Radial velocity dispersion of luminous
    satellites within $300$ kpc of each primary in the {\small APOSTLE} 
    runs, as a function of the host's virial mass. 
    The red crosses and blue plus signs correspond to the resolution levels, 
    L1 and L2, respectively. The thin dashed lines connect halos of 
    different resolutions. The solid
    coloured line indicates the best power-law fit
    $\sigma \propto M^{1/3}$; the shaded areas indicate the rms around
    the fit. Note the strong correlation between velocity dispersion
    and host mass; the relatively low radial velocity dispersion of
    the MW satellites is best accommodated with a fairly low virial
    mass, lower than expected from abundance matching (see upward
    pointing arrow, the same as in Fig.~\ref{FigNSats}). {\it Bottom:}
    Same as top, but for the line-of-sight velocity dispersion of one
    satellite system as seen from the other primary. Compared with the
    observed line-of-sight velocity dispersion of M31 satellites,
    shown by the horizontal line, this measure favours a lower virial
    mass than inferred from abundance-matching analyses, indicated by
    the arrow. }
\label{FigSigma}\ec
\end{figure}
\begin{figure}
  \bc \hspace{-0.2cm}
  \resizebox{8.cm}{!}{\includegraphics{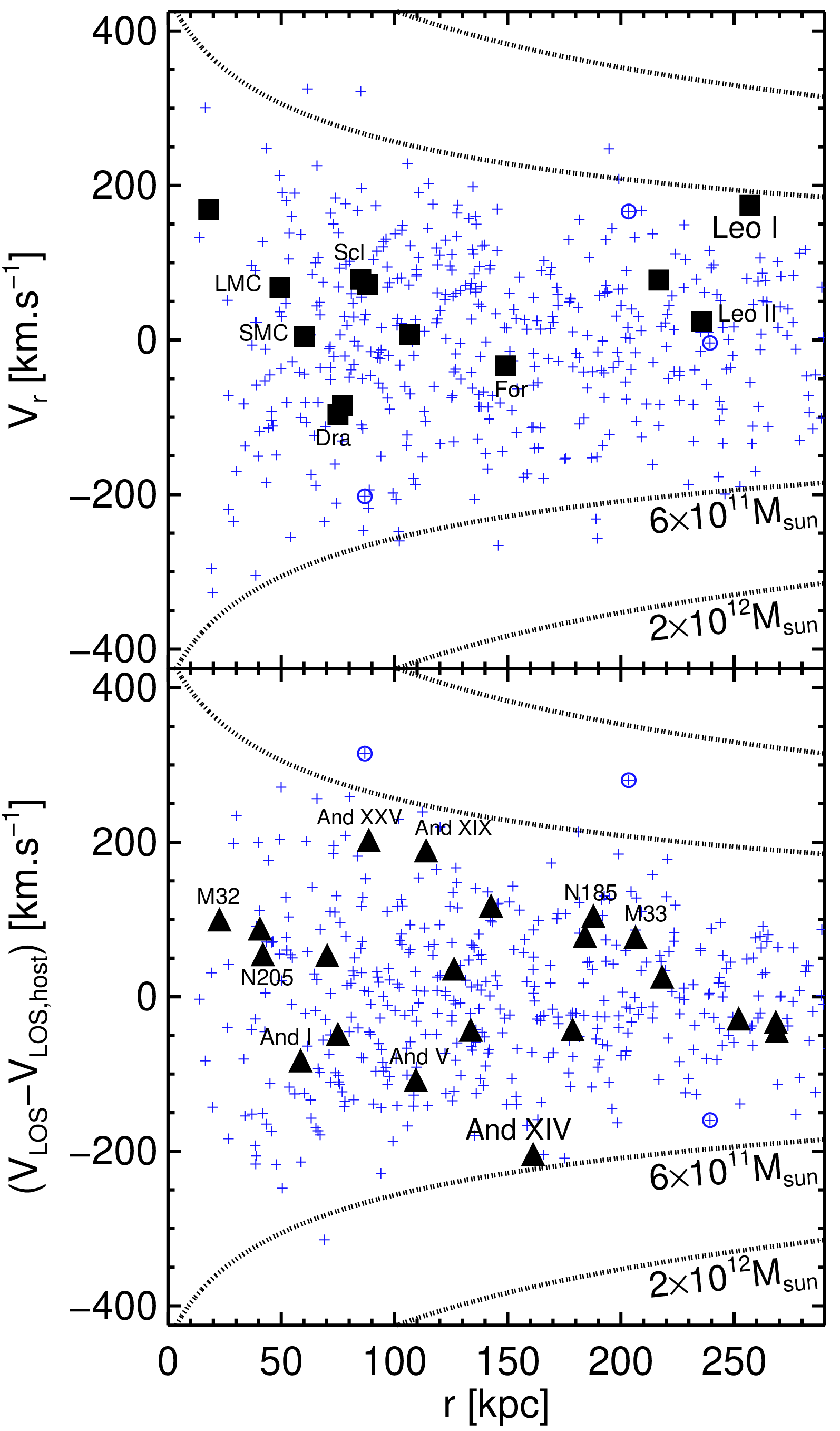}}\\%
  \caption{{\it Top}: Radial velocity vs distance for all satellites
    in the medium-resolution (L2) {\small APOSTLE} runs (``plus''
    symbols), compared with the observed radial velocities of the MW luminous
    satellites (filled squares). The positions of observed satellites in this
    phase-space correspond to regions that are well populated in the
    simulations. Very few satellites are ``unbound'' (marked by
    circles), as judged by escape velocity curves computed assuming a
    Navarro-Frenk-White profile of given virial mass (see dotted
    lines). {\it Bottom:} Same as top but for line-of-sight velocities
    of satellites as seen from the other primary. This may be directly
    compared with data for M31 satellites (filled triangles).}
\label{FigVRad}\ec
\end{figure}

\subsubsection{Most massive satellites}

Another concern when adopting a relatively low virial mass for the LG
primaries (compared to timing-argument and abundance-matching
  estimates), is that satellites as bright and massive as M33 in the
case of M31, or the LMC in the case of the MW, may fail to form. We
explore this in Fig.~\ref{FigBrightSats}, where we plot the stellar
mass of the primaries (open circles) and that of their most massive
satellite found within $300$ kpc (crosses, ``plus'', and square
symbols for L1, L2, and L3, respectively) as a function of their peak
circular velocity.  We also plot, for reference, the rotation velocity
and stellar masses of the MW and M31, as well as those of their
brightest satellites, the LMC and M33, respectively. For the LMC, the
error bar indicates the velocity range spanned by two different
estimates, $64\kms$ from \citet{vanderMarel2002}, and $87\kms$ from
\citet{Olsen2011}. For M33, we use a rotation speed of $110\kms$, with
an error bar that indicates the range of velocities observed between
$5$ and $15$ kpc by \citet{Corbelli2014}: we use this range as an
estimate of the uncertainty because the gaseous disk of M33 has a very
strong warp in the outer regions which hinders a proper determination
of its asymptotic circular speed.

Fig.~\ref{FigBrightSats} makes clear that there is no shortage of
massive satellites in the {\small APOSTLE} simulations: $6$ out
of $24$ primaries in the L2 runs have a massive subhalo with
$V_{\rm max}$ exceeding $100\kms$ (comparable to M33), and $12$ out of
$24$ have $V_{\rm max}>60\kms$ (comparable to the LMC). This result
is robust to numerical resolution; the numbers above change only to
$7$ and $11$, respectively, when considering the lower-resolution L3
runs (see open squares in Fig.~\ref{FigBrightSats}).

With hindsight, this result is not entirely surprising. According to
\citet{Wang2012a}, the {\it average} number of satellites scales as
$10.2\, (\nu/0.15)^{-3.11}$, where $\nu=V_{\rm max}/V_{200}$ is the
ratio between the maximum circular velocity of a subhalo and the
virial velocity of the primary halo. A halo of virial mass
$10^{12}\Msun$ has $V_{200}=145\kms$ , so we expect, on average, that
$1$ in $11$ of such halos should host a satellite as massive as M33
and $1$ in $2.3$ one like the LMC\footnote{These numbers assume a
  maximum circular velocity of $100\kms$ for M33 and $60\kms$ for the
  LMC.}. Our results seem quite consistent with this expectation.

A second point to note from Fig.~\ref{FigBrightSats} is that the
stellar mass of the primary galaxies are below recent estimates for
either the MW \citep[$5.2 \times 10^{10}\Msun$, according
to][]{Bovy2013} or M31 \citep[$1.03 \times 10^{11}\Msun$,
according to][]{Hammer2007}.  Indeed, none of our $24$ primaries have
masses exceeding $5\times 10^{10}\Msun$, a consequence of the low
virial masses of our selected LG pairs coupled with the relatively low
galaxy formation efficiency of the {\small EAGLE} ``Ref'' model  in
$\sim 10^{12}\Msun$ halos. This may be seen in
Fig.~\ref{FigBrightSats}, where the open circles lie
systematically below the grey lines, which show abundance matching
predictions taken from three recent papers. 

This issue has been discussed by \citet{Schaye2015}, and is reflected
in their fig.~4, which shows that the {\small EAGLE} ``Ref'' model we
use here underpredicts the number of galaxies with stellar masses a
few times $10^{10}\Msun$.  That same figure shows that the
opposite is true for smaller galaxies: the ``Ref'' model actually
overpredicts slightly but systematically the number of galaxies with
stellar masses a few times $10^8\Msun$.

Galaxy formation efficiency in our runs thus seems slightly too low in
MW-sized halos, and slightly too high in systems of much lower
mass. This slight mismatch in the galaxy mass-halo mass relation
manifests itself more clearly on LG scales, making it difficult to
match simultaneously the stellar masses of the MW and M31 as well as
that of their luminous satellites. Indeed, increasing the halo masses
of our LG candidates would lead to a better match to the stellar
masses of the primaries, but at the expense of overpredicting
the number of bright satellites (see, e.g., Fig.~\ref{FigNSats}),
unless the star formation and feedback parameters 
are recalibrated.  

The stellar masses of LG galaxies are therefore
a sensitive probe of the galaxy formation efficiency on
cosmological scales, and provide an important constraint on the
ability of cosmological codes to reproduce the observed galaxy
population.


\subsection{Satellite kinematics}
\label{SecSatKin}

We can also use the kinematics of MW and M31 satellites to gauge the
consistency of our results with the simulated satellite population. We
begin by considering, in the top panel of Fig.~\ref{FigSigma}, the
radial velocity dispersion, $\sigma_{r}$, of all satellites more
massive than $1.4\times 10^5\Msun$ (brighter than $M_V=-8$)
within $300$ kpc from either primary. We expect the satellite velocity
dispersion to scale as
$\sigma_{r}\propto V_{200} \propto M_{200}^{1/3}$, so the solid line
indicates the best fit with that slope to the data for our $24$
systems. (Symbols are as in Fig.~\ref{FigNSats}; blue $+$ symbols
indicate L2 resolution, red crosses correspond to L1 resolution.) This
scaling describes the correlation shown in Fig.~\ref{FigSigma} fairly
well, and its best fit suggests that a $10^{12}\Msun$ system
should host a satellite system with a radial velocity dispersion of 
$98\pm17\kms$. (The shaded area in Fig.~\ref{FigSigma}
indicates the rms scatter about the $M^{1/3}$ best fit.)  The observed
dispersion of the MW satellites ($86\kms$; shown by the dashed
horizontal line) thus suggests a mass in the range
$3.9\times10^{11}\Msun$ to $1.1\times 10^{12}\Msun$.  The
scatter, however, is large and these data alone can hardly be used to
rule out larger virial masses.

The bottom panel of Fig.~\ref{FigSigma} shows a similar analysis, but
applied to the satellites of M31. Since M31-centric radial
velocities are not directly available from
  observations\footnote{The M31-centric radial velocities can in
    principle be inferred from the data, but only by making
further assumptions; see. e.g., \citet{Karachentsev2006}}, we repeat the analysis for
our $24$ satellite systems using {\it projected} velocities, measured
along the line of sight from the other primary of each pair. (The
analysis uses only satellites within $300$ kpc from the 
galaxy's center.) The relation between the projected velocity
dispersion, $\sigma_{\rm Vlos}$, and virial mass is again reasonably well
described by the expected $M_{200}^{1/3}$ scaling. In this case, the
observed projected dispersion of $99\kms$ for the M31 satellites suggests a mass
in the range $7.6\times10^{11}\Msun$ to $2.1  \times 10^{12}\Msun$ 
although the scatter is again large.

The main conclusion to draw from Fig.~\ref{FigSigma} is that the
observed velocity dispersions of both the MW and M31 satellites are well
within the ranges found in our simulations. The relatively low mass of
our pairs thus does not seem to pose any problems reproducing the
kinematics of the LG satellite population. On the contrary, as was the case for
satellite numbers, the satellite kinematics would be difficult to
reconcile with much larger virial masses. For example, if the MW 
had a virial mass of $1.8 \times 10^{12}\Msun$, as suggested by
abundance-matching \citep{Guo2010}, then its satellite systems would, on average,
have a velocity dispersion of order $120\kms$, which would exceed the
$86\kms$  observed for the MW companions.

Finally, we check whether any of the MW or M31 satellites would be
unbound given the mass of the primaries chosen for our sample. We
explore this in Fig.~\ref{FigVRad}, where we plot the observed
velocity (radial in the case of the MW; line-of-sight in the case of
M31) of satellites as a function of their distance to the primary's
center (solid symbols). 

The dotted lines delineate the escape velocity as a function of
distance for halos with virial massses of $6\times 10^{11}$ and $2\times
10^{12}\Msun$, corresponding to roughly the minimum and maximum virial masses
of all primaries in our sample. The escape velocities assume a
Navarro-Frenk-White profile \citep{Navarro1996,Navarro1997}, with a
concentration $c=10$. ``Escapers'' (i.e., satellites with $3D$
velocities exceeding the escape speed for its primary) are shown by
circled symbols. These are rare; only $3$ of the $439$
$M_V<-8$  satellites examined are moving with velocities exceeding
the nominal NFW escape velocity of their halo \citep[see
also][]{Boylan-Kolchin2013}.

It is clear from this figure that, for our choice of masses, none of
the MW or M31 satellites would be unbound given 
their radial velocity. Indeed, even Leo I and And
XIV (the least bound satellites of the MW and M31, respectively) are
both within the bound region, and, furthermore, in a
region of phase space shared with many satellites in our {\small APOSTLE} sample. 
The kinematics of the satellite populations of
MW and M31 thus seems consistent with that of our simulated satellite populations.


\section{Summary and conclusions}
\label{SecConc}

We have analyzed the constraints placed on the mass of the Local Group
by the kinematics of the MW-M31 pair and of other LG members. We used
these constraints to guide the selection, from a large cosmological
simulation, of $12$ candidate environments for the {\small
  EAGLE-APOSTLE} project, a suite of hydrodynamical resimulations run
at various numerical resolution levels (reaching
$\sim10^{4}\Msun$ per gas particle at the highest level) and aimed at
studying the formation of galaxies in the local Universe.

{\small APOSTLE} uses the same code and star formation/feedback
subgrid modules developed for the {\small EAGLE} project, which yield,
in cosmologically-representative volumes, a galaxy stellar mass
function and average galaxy sizes in good agreement with
observations. This ensures that any success of our simulations in
reproducing Local Group-scale observations does not come at the
expense of subgrid module choices that might fail to reproduce the galaxy population
at large.  We also compare the simulated satellite populations of the
two main galaxies in the {\small APOSTLE} resimulations with the
observed satellite systems of M31 and MW to assess consistency with
observation.

Our main conclusions may be summarized as follows:

\begin{itemize}

\item{The kinematics of the MW-M31 pair and of other LG members are
    consistent with 
    a wide range of virial masses for the MW and
    M31. Compared with halo pairs selected from the Millennium
    Simulations, the relatively fast approach velocity of MW and M31
    favours a fairly large total mass, of order $5\times 10^{12}\Msun$. 
    On the other hand, the small tangential velocity and the
    small deceleration from the Hubble flow of outer LG members argue
    for a significantly smaller mass, of order $6\times 10^{11}\Msun$. Systems that satisfy the three
    criteria are rare---only $14$ are found in a $(137$ Mpc$)^3$
    volume---and span a wide range
    of masses, from $2.3\times 10^{11}$ to $6.1\times 10^{12}\Msun$, 
    with a median mass $\sim 2\times 10^{12} \Msun$. }

\item{Given the wide range of total masses allowed, the $12$ candidate
    pairs selected for resimulation in the {\small APOSTLE}
    project were chosen to loosely match the LG kinematic criteria and
    to span a relatively narrow range of masses (from
    $1.6\times 10^{12}$ to $3.6\times 10^{12}\Msun$, with a median
    mass of $2.3\times 10^{12}\Msun$). This enables us to explore the
    cosmic variance of our results at fixed mass, and, potentially, to
    scale them to other mass choices, if needed. }

\item{Large satellites such as LMC and M33 are fairly common around
    our simulated galaxies, although their total virial mass is well
    below that estimated from the timing argument.}

\item{The overall abundance of simulated satellites brighter than
  $M_V=-8$ is a strong function of the virial mass assumed for the
  LG primary galaxies in our simulations. The relatively few
  ($12$) such satellites around the MW suggests a fairly low mass
  ($\sim 6\times10^{11} \Msun$); the same argument suggests a mass for M31
  about twice as large ($\sim 1.2\times 10^{12}\Msun$).  }

\item{The velocity dispersions of simulated satellites are 
    consistent with those of the MW and M31. This diagnostic also
    suggests that virial masses much larger than those adopted for the
    {\small APOSTLE} project would be difficult to reconcile with
    the relatively low radial velocity dispersion observed for the MW
    satellite population, as well as with the projected velocity dispersion of
    the M31 satellite population. }

\item{The primary galaxies in the simulations are less massive than
    current estimates for the MW and M31. The most likely reason for
    this is an inaccuracy in the subgrid modelling of star formation
    and feedback and its dependence on halo mass.}

\end{itemize}

Our overall conclusion is that, despite some shortcomings, the {\small
  APOSTLE} simulation suite should prove a wonderful tool to
study the formation of the galaxies that populate our cosmic backyard.

\section{Acknowledgments}

We would like to thank Dr. Lydia Heck for her support through the
computational facility at the ICC. We are grateful to Alan McConnachie
for his help in interpreting the observational data, and also Richard
Bower for reading the manuscript and giving us useful feedback.  We
also would like to thank the anonymous referee for the constructive
report.  This research was supported by the National Science
Foundation under Grant No. PHY11-25915 and the hospitality of the
Kavli Insitute for Theoretical Physics at UC Santa Barbara This work
was supported in part by the Science and Technology Facilities Council
(grant number ST/F001166/1); European Research Council (grant numbers
GA 267291 ``Cosmiway'' and GA 278594 ``Gas Around Galaxies''); the
European Research Council under the European Union’s Seventh Framework
Programme (FP7/2007-2013), the National Science Foundation under Grant
No. PHYS-1066293, the Interuniversity Attraction Poles Programme of
the Belgian Science Policy Office [AP P7/08 CHARM] and the hospitality
of the Aspen Center for Physics. T. S. acknowledges the Marie-Curie
ITN CosmoComp.  RAC is a Royal Society University Research Fellow.
This work used the DiRAC Data Centric system at Durham University,
operated by the Institute for Computational Cosmology on behalf of the
STFC DiRAC HPC Facility (www.dirac.ac.uk), and resources provided by
WestGrid (www.westgrid.ca) and Compute Canada / Calcul Canada
(www.computecanada.ca). The DiRAC system is funded by BIS National
E-infrastructure capital grant ST/K00042X/1, STFC capital grants
ST/H008519/1 and ST/K00087X/1, STFC DiRAC Operations grant
ST/K003267/1, and Durham University. DiRAC is part of the National
E-Infrastructure.

\bibliographystyle{apj}
\bibliography{master}

\appendix
\section{Parameters of the Initial Conditions}
\label{SecIC}
The initial conditions for the {\small DOVE} and {\small APOSTLE}
simulations were generated from the {\small PANPHASIA} white-noise
field \citep{Jenkins2013} using second-order Lagrangian perturbation
theory \citep{Jenkins2010}.
The coordinates of the centers and the radii of the 
high resolution Lagrangian regions in the initial conditions, as well as 
the positions of the  MW and M31 analogs at $z=0$, for the
twelve {\small APOSTLE} volumes are given in Table~\ref{TabIC}.

\begin{table*}
  \caption{The positions of main halos at $z=0$ and parameters of 
    the high resolution Lagrangian regions of 
    the {\small APOSTLE} volumes in the initial conditions. The
    first column labels each volume. The next columns list the 
    ({\small X,Y,Z}) coordinates of each of the primaries at $z=0$. 
    The final four columns give the comoving coordinate centre and radius 
    of a sphere that contains the high resolution Lagrangian region in the 
    initial conditions, for each of the zoom initial conditions. 
    The phase descriptor for the {\small APOSTLE} runs is, in {\small PANPHASIA} format, 
    {[Panph1,L16,(31250,23438,39063),S12,CH1292987594,DOVE]}.  }
  \bc
 \begin {tabular}{| l *{6}{r} *{4}{r}  |}
\hline
Name  &  $X_{1}$  &  $Y_1$    &  $Z_1$   &   $X_2$  &  $Y_2$    &  $Z_2$  & $X_{l}$  &  $Y_l$  &  $Z_l$   &  $R_l$  \\
      & [Mpc]     & [Mpc]    &  [Mpc]   &   [Mpc]  &  [Mpc]    & [Mpc]   &  [Mpc]   & [Mpc]   & [Mpc]   & [Mpc]   \\
\hline
AP-1  & 19.326    & 40.284   &	46.508  &   18.917 &  39.725   &  47.001 &  26.5   &	39.1    &  39.0   & 7.9     \\
AP-2  & 28.798	  & 65.944   &	17.153  &   28.366 &  65.981   &  16.470 &  28.1   & 	60.2    & 18.4	  & 14.3    \\
AP-3  & 51.604    & 28.999   &	11.953  &   51.091 &  28.243   &  12.061 &  46.0   &	31.7    & 11.6    & 16.8     \\
AP-4  & 63.668    & 19.537   &	72.411  &   63.158 &  20.137   &  72.467 &  57.1   &	20.6	& 74.9	  & 8.4     \\
AP-5  & 42.716	  & 87.781   &	93.252  &   42.872 &  88.478   &  93.671 &  40.8   &	85.4    & 91.8    & 13.8     \\
AP-6  & 35.968    & 9.980    &	43.782  &   36.171 &   9.223   &  43.251 &  32.9   &	13.1    & 45.2    & 9.9    \\
AP-7  & 91.590    & 43.942   &	14.826  &   91.822 &  43.323   &  14.885 &  99.3   &	39.7    & 15.9    & 12.9   \\
AP-8  &  4.619    & 22.762   &  85.535  &   4.604  &  23.508   &  85.203 &  4.9    &	20.4    & 89.9    & 9.9   \\
AP-9  &  57.044   & 88.490   &	74.765  &  57.496  &  87.889   & 74.456  &  55.2   &	93.4    & 76.5	  & 7.9   \\
AP-10 & 61.949    & 24.232   &	98.305  &  61.867  & 24.925    & 98.124  &  62.5   & 	24.5    & 93.5    & 11.7   \\
AP-11 &  12.564   & 48.080   &	35.249  &  12.484  & 47.793    & 35.959  &  18.3   &	43.1	& 29.9    & 8.2    \\
AP-12 &  97.553   & 89.587   &	72.093  &   97.351 & 90.100    & 72.407  &  98.5   &	91.9    & 81.9    & 7.9   \\
\hline
\end{tabular}
\ec
\label{TabIC}
\end{table*}

\label{lastpage}

\end{document}